\documentclass[twocolumn,prl,showpacs]{revtex4} 

\usepackage{epsfig}
\begin{document}

\title{Super-Arrhenius slowing down in the metastable liquid of hard
spheres\\}

\author{M. Dzugutov}

\affiliation{Department for Numerical Analysis and Computer Science, 
Royal Institute of Technology, SE--100 44 Stockholm, Sweden}

\date{\today}

\begin{abstract}
\noindent
It is demonstrated that a straightforward extension of the Arrhenius
law accurately describes diffusion in the thermodynamically stable
liquid of hard spheres. A sharp negative deviation from this behaviour
is observed as the liquid is compressed beyond its stability limit,
which can be compared with super-Arrhenius slowing down in supercooled
liquids. It is suggested that the observed dynamical transition is
caused by extra entropy barriers arising in the free-energy
landscape. This interpretation is corroborated by the observation of
characteristic anomalies in the geometry of a particle trajectory.
\end{abstract}


\maketitle

Computer simulations of liquids using hard spheres (HS) remain a key
source of information about the most fundamental aspects of the liquid
state. The main intellectual attraction of these models is in their
purely geometric nature: all the observable properties of a HS liquid
can be reduced to a single thermodynamic quantity - entropy, a measure
of finite-range structural correlations \cite{hansen, mountain}. Thus,
mapping a phenomenon exhibited by liquids with continuous interactions
on a HS model yields its interpretation in terms of the statistical
geometry of sphere packing.

An important aspect of liquid behaviour addressed using HS models is
supercooled liquid dynamics \cite{yip, gotze, cohen, speedy}. This
notion refers to the complex of dynamical anomalies observed in a
liquid which remains in a metastable equilibrium below its melting
point. For the one-component HS system, the stable liquid domain is
bounded by the critical value of the packing fraction $\eta_c = 0.493$
\cite{speedy} ($\eta = \pi \rho \sigma^3 /6$, where $\rho$ is the
density and $\sigma $ is the HS diameter). Having been compressed
beyond $\eta = 0.54$, the HS liquid inevitably crystallizes
\cite{speedy}.

The most prominent indicator of the supercooled liquid regime, and, in
fact, its defining feature \cite{ediger, angell}, is the
characteristic behaviour of the relaxation dynamics which slows down
with cooling progressively faster than it can be inferred from the
Arrhenius prescription universally describing the temperature
variation of the transport coefficients in stable liquids.

In contrast to atomistic models with continuous interaction, the HS
systems lack the energy scale. Therefore, in order to compare the
non-Arrhenius liquid dynamics as observed in the former with the
respective behaviour of the metastable HS liquid, we have to resolve
the fundamental problem of finding a description of the {\it stable}
HS liquid dynamics that would be an adequate analog of the Arrhenius
law. In this capacity, Hildebrant-Batchinski relation \cite{batch}
connecting the diffusion coefficient with the free volume was
suggested \cite{woodcock}. The HS liquid diffusion \cite{woodcock}
follows this relation in the stable domain and deviates from it for
$\eta > \eta_c$ \cite{angwood}. However, this deviation is {\it
positive} and, therefore, cannot be regarded as an appropriate analog
of the super-Arrhenius slowing down in supercooled liquids.

In this Letter, it is demonstrated that an earlier suggested relation
between the diffusion coefficient and the thermodynamic entropy
\cite{nat} can be interpreted as a straightforward extension of the
Arrhenius law. Having been tested by the molecular dynamics, it is
found to accurately describe the diffusion in the stable one-component
HS liquid. Moreover, a significant slowing down as compared with this
relation is observed in the metastable HS liquid domain, a direct
analogue of the super-Arrhenius slowing down in supercooled
liquids. This transition in the HS dynamics is discussed in terms of
the free energy landscape.

The Arrhenius law asserts that the diffusion coefficient, $D$, scales
with the temperature $T$ as
\begin{equation} 
D = D_0 e^{- A/k_BT}
\end{equation} 
where $A$, the so-called activation energy, is interpreted as the
average height of the free-energy barrier, per particle, that has to
be crossed in order to perform an elementary step in the process of
diffusion. It can be presented as $A = E_a -k_B TS_a$, where $E_a$ and
$S_a$ represent, respectively, the heights of the energy barrier and
the entropy barrier involved. For the HS liquid, $A / k_B T = -S_a$;
$e^{S_a}$ can be understood as the inverse average number of attempts
per a successful diffusive transition of a particle. Two possible
types of the free-energy landscape can be discussed in the context:

(i) The available configurations are abundantly connected; the
connectivity is facilitated by independent motions of individual
particles, and each particle has an unrestricted immediate access to
the entire space volume assigned to it by the structural
constraints. Then the inverse average number of attempts needed for a
particle to move from its current position scales as $e^{-A/k_B T} =
e^s$ where $s$ is the excess entropy, per particle, - the difference
between the system's entropy and that of the perfect gas at the same
thermodynamic conditions.

(ii) The configuration-space connectivity is restricted by additional
constraints, complementary to the ensemble-averaged structural
constraints dominating type (i) landscape. In this case, the diffusive
motions of an individual particle are coupled, in a presumably
hierarchical manner, with the respective motions of other particles
within a certain range \cite{pwa}. These dynamical constraints give
rise to extra high entropy barriers which divide the configuration
space into a set of components \cite{palmer}. These barriers dominate
the relaxation dynamics and, in this way, control the diffusion
rate. In terms of Eq. (1) this means that $e^{-A/k_B T} < e^s$ It has
to be emphasized that this type of free-energy landscape exists on a
limited time-scale and, therefore, is quite distinct from the
equilibrium distribution. However, there exists a conjecture \cite{ag}
expressing the cooperativity range in terms of equilibrium entropy;
this conjecture is not discussed here.

The described landscape regimes can be understood by keeping in mind
an important topological peculiarity of the multidimensional
configuration space \cite{ma} where each point is adjacent to almost
all other points. This can be illustrated by the ``small world''
topology of abundantly connected multidimensional networks
\cite{sw}. On the other hand, if an entropy barrier characteristic of
type (ii) landscape arises, it will be immediately close to almost all
points of the configuration space with a profound effect on the
general rate of dynamics.
 
The central postulate we adopt here is that the free-energy landscape
that controls the relaxation dynamics in the stable HS liquid is of
type (i). Having assumed this, Arrhenius law (1) can be transformed
into the following simple form relating $s$ and $D$:
\begin{equation} 
D = D_0 e^{s}
\end{equation} 
If the time-scale is measured in terms of the Enskog collision
frequency $\Gamma$ \cite{chap}, this relation becomes equivalent to
the earlier suggested scaling law for atomic diffusion \cite{nat}
which was successfully tested on a number of simple liquids, including
the HS liquid using a two-body approximation for $s$
\cite{mountain}. In the full $s$ version, the scaling law was tested
on some metallic liquids \cite{babak}, but for the HS liquid, such
tests have not been done so far. In this context, a conjecture should
be mentioned where $D \propto e^{Bs}$ with $B \neq 1$ \cite{rosen}.

In this study, relation (2) was tested using a molecular-dynamics
simulation of a HS system comprising 6912 identical particles. This
simulation was necessitated by the fact that, although the HS
diffusion has been explored in a number of simulations, a considerable
uncertainty remains concerning its size-dependence at different
densities \cite{speedy, erpenbeck}.

The assumption that the dynamics in a liquid is controlled by the
excess entropy, a measure of the structural correlations, implies that
the length should be measured in terms of the characteristic length of
the structure which manifests itself in the position of the main peak
of the structure factor. Fig. 1 shows that the latter scales with the
density $\rho$ as $\rho^{1/3}$; therefore, $\rho^{-1/3}$ is a
convenient unit of length. Fig. 1 also shows that this unit of length
is consistent with the equilibrium separation of the nearest
neighbours in the Lennard-Jones liquid of which the structure factor
can be reproduced by that of the HS liquid. The use of $\sigma$ as the
unit of length that was adopted in the earlier test of the scaling law
\cite{nat} on the HS liquid is apparently inappropriate at lower
densities.

\begin{figure}
\centerline{\epsfig{file=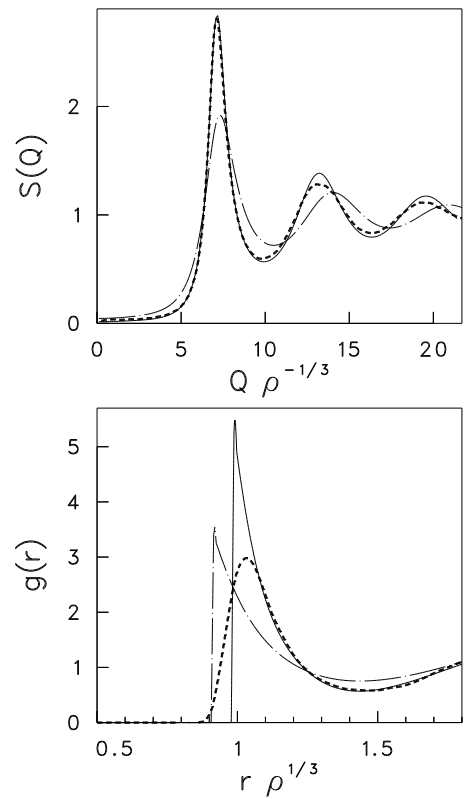,width=6.2cm,clip=}}
\caption{ The structure factor and the radial distribution function of
the HS liquid. Solid line, $\eta = 0.49$; dash-dotted line, $\eta =
0.38$. The dashed line, in both plots, the Lennard-Jones liquid
simulated at $\rho=0.85$ and $T=0.7$. }
\label{fig1}
\end{figure}

The HS excess entropy was calculated here by the thermodynamic
integration using Boublik-Nezbeda approximation \cite{boublik} for the
equation of state. A minor correction was introduced for $\eta>0.47$
\cite{speedy}.

The simulation results are shown in Fig. 2. Each value of $D$ was
obtained from the mean-square displacement averaged over $10^4$
collisions per particle. These results are compared with another set
of the HS diffusion data which represent an extrapolation to the
infinite system size\cite{erpenbeck, alder}. It is clear that with
$D_0 = 0.079$, relation (2) accurately describes the diffusion in the
one-component HS system within the range $0.36<\eta<0.49$ which covers
the entire stable liquid domain.

$D_0$ corresponds to the diffusion rate in a hypothetical situation
where the discussed liquid diffusion mechanism is still valid but
$s=0$. The observed value of $D_0$ can be rationalized using the
following arguments based on a simple toy model in the spirit of the
lattice gas models. Consider a tessellation of the space into
equal-size cells. Each cell contains one stationary particle which
stays within the cell until its status is changed; besides, there is
a number of mobile particles distributed in space. At each time-step,
a mobile particle collides with the stationary particle that is in the
same cell. As a result, the former becomes stationary, whereas the
latter becomes mobile and traverse the cell boundary into a randomly
chosen adjacent cell (if several mobile particles simultaneously occur
in one cell, an arbitrary collision sequence is assumed). Thus, each
particle performs a random walk with the step size $\approx
\rho^{-1/3}$. Assuming that the number of mobile particles is small,
and keeping in mind that there is one particle jump per two
collisions, the diffusion coefficient can be estimated as $D \approx
\Gamma \rho^{-2/3} /12$, which is in a good agreement with the above
value of $D_0$.
\begin{figure}
\centerline{\epsfig{file=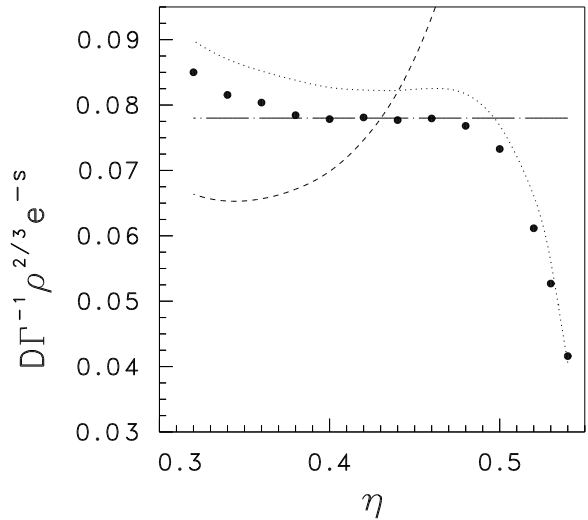,width=7.5cm,angle=-90}}
\caption{ Diffusion coefficient in the HS liquid as a function of the
packing fraction $\eta$. Dots, the present molecular dynamics
simulation; dotted line, the extrapolated limit for the infinite
system size \cite{erpenbeck, alder}; chain-dashed line: Eq. (2)XS with
$D_0=0.078$. Dashed line: the Enskog approximation \cite{chap}.}
\label{fig2}
\end{figure}

\begin{figure}
\centerline{\epsfig{file=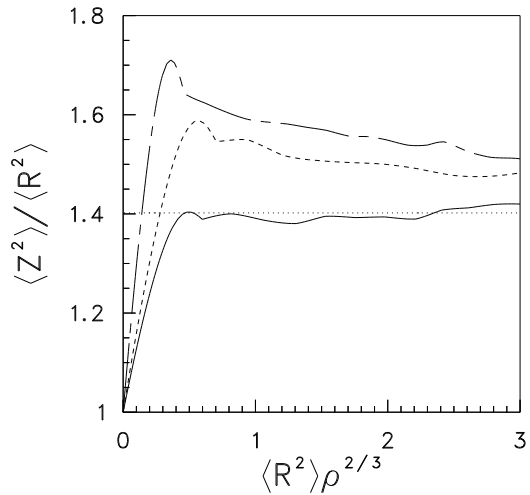,width=7cm,angle=0}}
\caption{The ratio of the mean square maximum displacement $\langle
Z^2(t) \rangle $ to the mean square displacement $\langle R^2(t)
\rangle $. Solid line, $\eta=0.48$; dashed line, $\eta=0.52$;
chain-dashed line, $\eta=0.53$.  Dotted line indicates the random walk
limit $\langle Z^2(t) \rangle / \langle R^2(t) \rangle = 1.4024...$
\cite{seshadri}.}
\label{fig3}
\end{figure}

The results presented in fig. 2 demonstrate that in the metastable
liquid domain $\eta>\eta_c$, the HS diffusion exhibits a rapidly
increasing slowing down as compared with the Arrhenius-like behaviour
conjectured by relation (2). Apparently, this behaviour can be
regarded as a direct analog of the super-Arrhenius slowing down in
the supercooled liquids with continuous interactions. Using the above
arguments, the onset of this new dynamical regime can be interpreted
as a clear signal that the height of entropy barriers exceeds the
ensemble-averaged excess entropy, and the free-energy landscape that
controls the HS liquid dynamics transforms from type (i) to type (ii).

The long-time configuration-space constraints conjectured to dominate
the HS liquid dynamics in the metastable domain can be detected from
their impact on a particle trajectory. The latter represents a 3D real
space projection of the multidimensional trajectory of the system in
its configurational space. The relaxation dynamics that unfolds in
type (i) landscape represents a random walk in the configurational
space, which must also be true for a single particle trajectory. The
long-time constraints caused by the development of extra barriers
characteristic of the type (ii) landscape give rise to distinct
anomalies in the geometry of a particle trajectory. A quantity that
can be conveniently employed as an indicator of a possible deviation
from the random walk geometry in a particle trajectory is the maximum
absolute value of the displacement from the original position within
the time interval $(0,t)$:
\begin{equation} 
Z(t) \equiv max \{  R(\tau ) , \hspace{.15in} 0 < \tau < t \} 
\end{equation} 
The ratio of its second moment $\langle Z^2(t) \rangle$, the mean
square maximum displacement, to the mean square displacement $\langle
R^2(t) \rangle$ for a random walk in 3D space is equal to $1.4024... $
\cite{seshadri}. The evolution of this ratio as a function of $\langle
R^2(t) \rangle$ for various values of $\eta$ is presented in
Fig. 3. For $\eta=0.48$, in the stable liquid domain, the random-walk
limit is attained after a short initial period of ballistic behaviour,
as soon as a particle leaves the cage of its nearest neighbours. The
geometry of a particle trajectory changes significantly as the HS
liquid is compressed beyond its stability limit: a new regime of
diffusion develops where $\langle Z^2(t) \rangle / \langle R^2(t)
\rangle$ exceeds the random walk value. This indicates that a particle
trajectory has a higher chance than random to return to an earlier
covered region. The extent of this apparent confinement effect in a
particle trajectory amounts to the diffusive displacement of several
particle diameters, and it increases rapidly, as well as the magnitude
of the effect, with the increase of $\eta$.

According to the arguments presented above, the height of an entropy
barrier can be regarded as a measure of the real-space extent of the
local configurational transformation that the system has to perform to
cross the barrier. Extra high barriers in type (ii) landscape
correspond to extended correlations, both positional and
dynamical. The rapid increase in the (time-limited) correlation length
in the metastable HS liquid that has been detected here from the
geometry of a particle trajectory represents a generic feature of the
supercooled liquid behaviour \cite{fischer}. Ergodicity restoring
relaxation whereby the extra high entropy barriers are created and
destroyed is facilitated by highly collective activated hopping. This
dynamics was discerned in the metastable HS liquid using dynamical
density-functional theory analysis \cite{kawasaki}.

The profound similarity between the pattern of slowing down in the
metastable HS liquid and that in conventional supercooled liquids
indicates the entropic nature of this phenomenon \cite{ritort}, and
suggests a unifying scenario for its development in terms of the
free-energy landscape transformation. The change in the packing
geometry under cooling/compression beyond the liquid phase stability
limit, presumably caused by reduction of the free volume, leads to
formation of long-lived extra high entropy barriers and the loss of
the short time-scale configuration-space connectivity. This results in
changing the relation between between $s$ and $D$ which can be
detected as a deviation from the scaling law (2).

This study was supported by the Swedish Research Council.


\begin{thebibliography}{199}
\bibitem{hansen} J. P. Hansen and I. McDonald, {\it Theory of Simple
Liquids}, (Academic Press, London, 1976)
\bibitem{mountain} R. D. Mountain, and H. Raveche,
Journ. Chem. Phys. {\bf 35}, 2250-2255 (1971)
\bibitem{yip} J. P. Boon and S. Yip, {\it Molecular Hydrodynamics},
McGraw-Hill, New York (1980)
\bibitem{gotze} W. G\"otze, and L. Sj\"ogren, Rep. Progr.  Phys., {\bf
55}, 241 (1992)
\bibitem{cohen} E. D. G. Cohen, Physica A, {\bf 194 }, 229-257 (1993)\\
 
\bibitem{speedy} R. J. Speedy, Mol. Phys., {\bf 95}, 168 (1998)
\bibitem{ediger} M. D. Ediger, C. A. Angell, and S. R. Nagel,
J. Phys. Chem., {\bf 100}, 13200 (1996)
\bibitem{angell} C. A. Angell, Journ. of Non-Cryst. Solids, {\bf
131-133}, 13 (1991)  
\bibitem{batch} A.J. Batchinski, Z. Phys. Chem., {\bf 84}, 643 (1913);
J.H. Hildebrand, {\it Viscosity and Diffusion}, (Wiley, New York,
1977)
\bibitem{woodcock}L.W. Woodcock, J. Chem. Soc. Faraday Trans. 2,
{\bf 72}, 1667 (1976)
\bibitem{angwood} L.W. Woodcock and C.A. Angell, Phys. Rev. Lett.,
{\bf 47}, 1129 (1981)
\bibitem{nat} M. Dzugutov, Nature, {\bf381}, 137-139 (1996)
\bibitem{pwa} R. G. Palmer, D.L. Stein, E. Abrahams, and
P.W. Anderson, Phys. Rev. Lett. {\bf 53}, 958 (1984)
\bibitem{palmer} R. G. Palmer, Adv. in Phys. {\bf 31}, 669 (1982)
\bibitem{ag} G. Adam and J. H. Gibbs, Journ. Chem. Phys. {\bf
43}, 139 (1965)
\bibitem{ma}  S.-K. Ma,  { \it Statistical mechanics}, World
Scientific, Singapore (1985)
\bibitem{sw} D. J. Watts and S. H. Strogatz, Nature, {\bf 393}, 440
(1998); R.Albert, H. Jeong, and A.-L. Barabasi, Nature, {\bf 406}, 378
(2000)
\bibitem{chap} S. Chapman and T. G. Cowling, {\it The mathematical
theory of non-uniform gases}, (University Press, Cambridge, 1939)
\bibitem{babak} J. Hoyt, M. Asta, and B. Sadigh, Phys. Rev. Lett.,
{\bf 85}, 594 (2000)
\bibitem{rosen} Y. Rosenfeld, Phys. Rev. A {\bf 15}, 2545 (1977);
Y. Rosenfeld, J. Ppys.; Cond. Matt., {\bf 11}, 5415 (1999)
\bibitem{erpenbeck} J. J. Erpenbeck, W. W. Wood, Phys. Rev. A,
{\bf 43}, 4254, (1991)

\bibitem{alder} B.J.Alder, D.M.Gass, and T.E.Wainwright,
J. Chem. Phys., {\bf 53}, 3813, (1991)
\bibitem{boublik} T. Boublik and J. Nezbeda,
Coll. Czech. Chem. Commun., {\bf 51}, 2301 (1985)
\bibitem{seshadri} V. Seshadri and K. Lindenberg, J. Stat. Phys., {\bf
22}, 69 (1980)
\bibitem{fischer} E. W. Fischer, E. Donth, and W. Steffen,
Phys. Rev. Lett. {\bf 68}, 2344 (1992); W. Kob, C. Donati,
S. J. Plimpton, P. H. Pool, and S. C. Glotzer, Phys. Rev. Lett. {\bf
79}, 2827 (1997); E. R. Weeks, J. C. Crocker, A. C. Levitt,
A. Scofield, and D. A. Weitz, Science, {\bf 287}, 627 (2000)
\bibitem{kawasaki} K. Fuchizaki, K. Kawasaki, J. Phys. Soc. Japan,
{\bf 67}, 2158, (1998)
\bibitem{ritort} A. Crisanti and F. Ritort, Cond-mat/0102104 (2001)




\end{thebibliography}
\end{document}